\documentclass[journal]{IEEEtran}
\ifCLASSINFOpdf
\else
   \usepackage[dvips]{graphicx}
\fi
\usepackage{url}

\usepackage{graphicx}

\usepackage{amsmath,amssymb,threeparttable,placeins,makecell,stfloats,cite}
\usepackage[usenames]{color}
\usepackage{amsfonts}
\usepackage{latexsym}
\usepackage{lipsum}
\usepackage{floatrow}
\newtheorem{theorem}{{{\textit{Theorem}}}}

\newtheorem{lemma}{{{\textit{Lemma}}}}

\newtheorem{definition}{{{\textit{Definition}}}}

\newtheorem{remark}{{{\textit{Remark}}}}

\newtheorem{example}{{{\textit{Example}}}}

\begin{document}
\title{Pseudo-Boolean Functions for Optimal Z-Complementary Code Sets with Flexible Lengths}
\author{Palash~Sarkar,~
        Sudhan~Majhi,~
        and~Zilong~Liu

\thanks{Palash Sarkar and Sudhan Majhi are with the Department of EE, IIT Patna, India, e-mail: {\tt\{ palash.pma15,smajhi\}}@iitp.ac.in.}
\thanks{Zilong Liu is with the School of CSEE, University of Essex, UK, e-mail:{\tt zilong.liu@essex.ac.uk}.}}
\IEEEpeerreviewmaketitle
\maketitle
\begin{abstract}
This paper aims to construct optimal Z-complementary code set (ZCCS) with non-power-of-two (NPT) lengths to enable interference-free multicarrier 
code-division multiple access (MC-CDMA) systems. 
The existing ZCCSs with NPT lengths, which are constructed from generalized Boolean functions (GBFs), 
are sub-optimal only with respect to the set size upper bound.   
For the first time in the literature, we advocate the use of pseudo-Boolean functions (PBFs)
(each of which transforms a number of binary variables to a real number as a natural generalization of GBF) for direct 
constructions of optimal ZCCSs with NPT lengths.
\end{abstract}
\section{Introduction}\label{sec:intro}
\IEEEPARstart{M}{ulticarrier} code-division multiple access (MC-CDMA) has been one of the most widely adopted wireless techniques in many communication 
systems/standards owing to its efficient fast Fourier transform (FFT) based implementation, 
resilience against intersymbol interference, and high spectral efficiency \cite{chen2007next}.
That being said, MC-CDMA may suffer from multiple-access
interference (MAI) \cite{mai1} and multipath interference (MPI) \cite{mpi1}. 
A promising way to address both MAI and MPI is to adopt proper spreading codes, such as
complete complementary codes (CCC) \cite{arthina} and Z-complementary code sets (ZCCSs) \cite{psktcom}.
This paper focuses on efficient construction of ZCCSs with a new tool, called pseudo-Boolean functions (PBFs), to enable interference-free 
quasi-synchronous MC-CDMA systems.

In 2007, Z-complementary pairs (ZCPs) were introduced by  Fan \emph{et~al.} \cite{fan2007} to overcome the limitation on the lengths of Golay complementary pairs (GCPs) \cite{gol1961,Davis1999}. 
A ZCP refers to a pair of sequences of same length $N$ having zero aperiodic auto-correlation sums for all
time shifts $\tau$ satisfying $0<|\tau|<Z$, where $Z$ is called zero-correlation zone (ZCZ) width. When $Z=N$, the resultant sequence pair reduces to 
a GCP. 
In the literature, there are direct constructions of GCPs and ZCPs with the aid of generalized Boolean functions (GBFs) \cite{chen,czcp3,wespl}.
The idea of ZCPs introduced
in \cite{fan2007} was generalized to ZCCS by Feng \emph{et~al.} in \cite{lfengispl2008}. A ZCCS refers to a set of
$K$ codes, each of which consists of $M$ constituent sequences of identical length $L$, having ideal aperiodic auto- and cross-correlation properties inside 
the ZCZ width and satisfying
the theoretical upper bound: $K\leq M\lfloor N/Z\rfloor$ \cite{liu2011}. When $Z=N$, the set is called a mutually orthogonal Golay complementary sets (MOGCSs) \cite{arthina},
which refers to collection of complementary codes (CCs) \cite{pater2000,isit19,tcom20} with ideal cross-correlation properties.
A set of CCCs is known as a MOGCSs with the equality $K=M$ \cite{cccsmajhi,uda2014,Liu_FDRR_2015,slett}. The theoretical upper bound shows that an optimal 
ZCCS has larger set size as compared to CCCs provided $\lfloor \frac{N}{Z}\rfloor\geq 2$. Recently, several GBFs based
constructions of optimal ZCCSs with power-of-two lengths have been reported in \cite{psktcom,ps2cd,chuzccs,chenzcom}. In the recent literature, two direct constructions of ZCCSs with NPT lengths can be found in   
\cite{palnpl} and \cite{chuzccs}, which produces sub-optimal ZCCS with $\lfloor \frac{N}{Z}\rfloor=1$ and non-optimal ZCCSs for NPT lengths
with $\lfloor \frac{N}{Z}\rfloor<1$, respectively.
To the best of our knowledge, the construction of optimal ZCCSs of NPT lengths
with $\lfloor \frac{N}{Z}\rfloor\geq 2$, based on GBFs remains open. Other methods which are dependent on the existence of special sequences,
known as indirect constructions \cite{wespl}, to construct ZCCSs can be found in
\cite{tibudazccs,avezccs,jli_igc_2008}. The indirect constructions heavily rely on a
series of sequence operations which  may not be
feasible for rapid hardware generation, especially, when the
sequence lengths are large \cite{psktcom}.
\begin{table}[!t]
\centering
\caption{Comparison of the Proposed Construction with \cite{psktcom,chenzcom,palnpl,ps2cd,chuzccs,jli_igc_2008} }\label{ointa}
\resizebox{\textwidth}{!}{
\begin{tabular}{|l|l|l|l|l|l|l|}
\hline
ZCCS           & Method       &Length ($N$)& $\lfloor\frac{N}{Z}\rfloor$&Constraints     &Optimality \\ \hline
\cite{psktcom} &Direct        &$2^m$     & $\geq 2$                     &$m\geq 2$       &Optimal     \\ \hline
\cite{chenzcom}&Direct         &$2^m$     &$\geq 2$                      &$m\geq 2$       &Optimal     \\ \hline
\cite{palnpl}  &Direct         &$2^m+2$   &$=1$                          & $m\geq 4$      &Sub-optimal     \\ \hline
\cite{ps2cd}   &Direct          &$2^m$     &$\geq 2$                      &$m\geq 2$       &Optimal          \\ \hline
\cite{chuzccs} &Direct         &$2^m$     &$\geq 2$                      &$m\geq 2$       &Optimal           \\ \hline
\cite{chuzccs} &Direct        &$2^m+2^h$ &$\geq 1$                         & $m>0,0<h\leq m$&Non-optimal        \\ \hline
\cite{jli_igc_2008}&Indirect   &     $L$        &$\geq 2$                      &$L\geq 1$ & Optimal            \\ \hline
\textit{Theorem 1} &Direct          &$p2^m$    &$\geq 2$                      &$m\geq 2$, prime $p$   &Optimal     \\ \hline
\end{tabular}}
\end{table}

It is noted that 
the MAI in MC-CDMA system can be mitigated using zero-correlation properties of a ZCCS provided that all the received multiuser signals are roughly synchronous within the ZCZ width \cite{Liu_FDRR_2015}. 
In addition to their applications in
MC-CDMA \cite{uda2014,Liu_FDRR_2015,jli_igc_2008}, ZCCSs have also been employed as optimal training sequences in multiple-input multiple-output (MIMO)
communications \cite{fan2008,hmwang2007}. The limitation on the set size of CCCs and the unavailability of optimal ZCCSs with 
NPT lengths using direct constructions in the existing literature are a major motivation of this work. Specifically, for the first time in the literature, we propose to use PBFs for direct construction of optimal 
ZCCS of lengths $p2^m$, where $p$ is a prime number and $m$ is a positive integer. A PBF \cite{psbb} refers to an arbitrary mapping of the set 
of binary $m$-tuples to real numbers. Being a natural generalization of GBFs, PBFs are also suitable for rapid hardware 
generation of sequences.
A detailed comparison of the proposed construction with \cite{psktcom,chenzcom,palnpl,ps2cd,chuzccs,jli_igc_2008} is given in TABLE I.
%
\section{Preliminary}
In this section, we present some basic definitions and lemmas to be used in the proposed construction.
Let $\textbf{y}_1=(y_{1,0},y_{1,1},\cdots, y_{1,N-1})$ and $\textbf{y}_2=(y_{2,0},y_{2,1},\cdots, y_{2,N-1})$ denote a pair of sequences with complex components. 
For an integer $\tau$, define \cite{psktcom}
\begin{equation}\label{equ:cross}
\theta(\textbf{y}_1, \textbf{y}_2)(\tau)=\begin{cases}
\sum_{i=0}^{N-1-\tau}y_{1,i+\tau}y^{*}_{2,i}, & 0 \leq \tau < N, \\
\sum_{i=0}^{N+\tau -1}y_{1,i}y^{*}_{2,i-\tau}, & -N< \tau < 0,  \\
0, & \text{otherwise},
\end{cases}
\end{equation}
The functions $\theta(\textbf{y}_1,\textbf{y}_2)$ and $\theta(\textbf{y}_1,\textbf{y}_1)$ are called the aperiodic cross-correlation function (ACCF)
between $\textbf{y}_1$ and $\textbf{y}_2$, and the aperiodic auto-correlation function (AACF) of $\textbf{y}_1$, respectively.
Let $\mathcal{S}=\{\mathcal{S}_0,\mathcal{S}_1, \cdots ,\mathcal{S}_{K-1}\}$ be a set of $K$ codes or ordered sets
defined as
\begin{equation}
\begin{split}
\mathcal{S}_\mu=
\left(\textbf{s}_0^\mu,\textbf{s}_1^\mu,\hdots,\textbf{s}_{M-1}^{\mu}\right),
\end{split}
\end{equation}
where $\textbf{s}_\nu^\mu$ $(0\leq \nu \leq M-1,0 \leq \mu \leq K-1)$ is the $\nu$-th element which we assume is a complex-valued sequence of length
$N$ in $\mathcal{S}_\mu$.
For $\mathcal{S}_{\mu_1}$, $\mathcal{S}_{\mu_2}\in \mathcal{S}$ $(0\leq \mu_1,\mu_2\leq K-1)$, the ACCF
between $\mathcal{S}_{\mu_1}$ and $\mathcal{S}_{\mu_2}$ is defined as
 \begin{equation}
  \theta(\mathcal{S}_{\mu_1},\mathcal{S}_{\mu_2})(\tau)=\displaystyle \sum_{\nu=0}^{M-1}\theta(\textbf{s}_\nu^{\mu_1},\textbf{s}_\nu^{\mu_2})(\tau).
 \end{equation}
\begin{definition}[\cite{psktcom}]
Code set $\mathcal{S}$ is called a ZCCS if
\begin{eqnarray}
\theta(\mathcal{S}_{\mu_1},\mathcal{S}_{\mu_2})(\tau)
=\begin{cases}
MN, & \tau=0, \mu_1=\mu_2,\\
0, & 0<|\tau|<Z, \mu_1=\mu_2,\\
0, & |\tau|< Z, \mu_1\neq \mu_2,
\end{cases}
\end{eqnarray}
where $Z$ is called ZCZ width. We denote a ZCCS with the parameters $K,N,M$, and $Z$ by the notation $(K,Z)$-$\text{ZCCS}_M^N$. 
For $K=M$ and $Z=N$, a $(K,Z)$-$\text{ZCCS}_M^N$ is called a set of CCCs and we denote it by $(K,K,N)$-CCC. 
\end{definition}
We call a $(K,Z)$-$\text{ZCCS}_M^N$ optimal if it achieves the equality in the theoretical upper-bound, given 
by $ K\leq M\left\lfloor \frac{N}{Z}\right\rfloor$ \cite{liu2011}.
\subsection{Generalized Boolean Functions (GBFs)}
Let $x_0,x_1,\hdots,x_{m-1}$ denote $m$ variables which take values from $\mathbb{Z}_2$. A monomial of degree $i$ ($0\leq i \leq m$) is defined as the product
of any $i$ distinct variables among  $x_0,x_1,\hdots,x_{m-1}$. Let us assume that $\mathcal{A}_i$ denotes the set of all monomials of degree $i$,
where 
\begin{equation}
 \begin{split}
  \mathcal{A}_i=&\left\{x_0^{r_0}x_1^{r_1}\cdots x_{m-1}^{r_{m-1}}:\right.\\&\left.r_0+r_1+\cdots+r_{m-1}=i,(r_0,r_1,\hdots,r_{m-1})\in\mathbb{Z}_2^m\right\}.
 \end{split}
\end{equation}
A function $f:\mathbb{Z}_2^m\rightarrow\mathbb{Z}_q$ is said to be a GBF if it can uniquely be expressed as a linear
combination of the monomials in $\mathcal{A}_m$, where the coefficient of each monomial is drawn from $\mathbb{Z}_q$, where $\mathbb{Z}_q$ denotes the set of integers modulo $q$. The highest degree monomial
with non-zero coefficient present in the expression of $f$ determine the order of $f$. As an example, $2x_0x_1+x_1+1$ is a second order GBF of two variables
$x_0$ and $x_1$ over $\mathbb{Z}_3$. We denote the graph of a second-order GBF $f$ by $G(f)$ \cite{pater2000}. It contains $m$ vertices which are denoted by the $m$ variables of $f$. 
The edges in the $G(f)$ are determined by the second-degree monomials present in the expression of $f$ with non-zero coefficients. 
There is an edge of weight $w$ between the vertices $x_\alpha$ and $x_\beta$ of $G(f)$ if the expression of $f$ contains the term $wx_\alpha x_\beta$. 
Let $\psi(f)$ denotes the complex-valued sequence corresponding to a GBF $f$ and it is defined as \cite{pater2000}, 
$\psi(f)=(\omega_q^{f_0}, \omega_q^{f_1}, \hdots, \omega_q^{f_{2^m-1}})$, where $\omega_q$ denotes $\exp\left(2\pi\sqrt{-1}/q\right)$, $f_r=f(r_0,r_1,\hdots,r_{m-1})$, 
$(r_0,r_1,\hdots,r_{m-1})$ is the binary vector representation of integer $r$ ($r=\displaystyle \sum_{\alpha=0}^{m-1}r_\alpha 2^\alpha$), and $q$ denotes an even number, no less than 2. 
We denote by $\bar{x}=1-x$
the binary complement of $x\in \{0,1\}$.
For any given GBF $f$ in $m$ variables, we denote the function $f(1-x_0,1-x_1,\hdots,1-x_{m-1})$ or $f(\bar{x}_0,\bar{x}_1,\hdots,\bar{x}_{m-1})$  by $\tilde{f}$.
Let $\mathcal{C}=\left(g_1,g_2,\hdots,g_{M}\right)$ be an ordered set of $M$ GBFs. We define the code $\psi(\mathcal{C})$ corresponding to 
$\mathcal{C}$ as $\psi(\mathcal{C})=\left(\psi(g_1),\psi(g_2),\hdots,\psi(g_{M})\right)$.
\begin{lemma}{(Construction of CCC \cite{arthina})}\label{lemma3}\\
Let $f:\mathbb{Z}_2^m\rightarrow \mathbb{Z}_q$ be a second-order GBF. Let us assume that $G(f)$
contains the vertices $x_{j_0},x_{j_1},\hdots,x_{j_{k-1}}$ such a way that 
after performing a deletion operation on those vertices, the resulting graph reduces to a path. Let the edges in the path have identical weight 
of $\frac{q}{2}$ and
$\mathbf{t}=(t_0,t_1,\cdots, t_{k-1})$ be the binary representation of the integer $t$. Define the CC, $C_t$ to be
\begin{equation}
 \Big\{ f\!+\!
 \frac{q}{2}\Big((\mathbf{d}+\mathbf{t})\cdot \mathbf{x}
 \!+\!dx_{\gamma}\Big): 
 \mathbf{d}\in \{0,1\}^k,d\in\{0,1\}     \Big\},
\end{equation}
and $\bar{C}_{t}$ to be
\begin{equation}
 \Big\{ \tilde{f}\!+\!
 \frac{q}{2}\Big((\mathbf{d}+\mathbf{t})\cdot \bar{\mathbf{x}}
 \!+\!\bar{d}x_{\gamma}\Big): \mathbf{d}\in \{0,1\}^k,d\in\{0,1\}\Big\},
 \end{equation}
where $(\cdot)\cdot(\cdot)$ denotes the dot product between two real-valued vector $(\cdot)$ and $(\cdot)$, $\gamma$ is the label of either end vertex in the path, $\mathbf{x}=(x_{j_0},x_{j_1},\hdots,x_{j_{k-1}})$, 
 $\bar{\mathbf{x}}=(1-x_{j_0},1-x_{j_1},\hdots,1-x_{j_{k-1}})$, and $\mathbf{d}=(d_0,d_1,\hdots,d_{k-1})$.
 Then
  $\{\psi(C_t),\psi^*(\bar{C}_{t}):0\leq t<2^k\}$ forms
$(2^{k+1},2^{k+1},2^m)$-CCC, where $\psi^*(\cdot)$ denotes the complex conjugate of $\psi(\cdot)$.
\end{lemma}
\subsection{Pseudo-Boolean Functions (PBFs)}
A function $F:\{0,1\}^m\rightarrow\mathbb{R}$ is said to be a PBF if it can be uniquely expressed as a linear
combination of monomials in $\mathcal{A}_m$ with the coefficients drawn from $\mathbb{R}$, where $\mathbb{R}$ denotes the set 
of real numbers. Therefore, PBFs are a natural generalization
of GBFs \cite{psbb}. As an example, $\frac{2}{3}x_0x_1+x_0+1$ is a second-order PBF of two variables $x_0$ and $x_1$ but not a GBF. 
Let $f:\mathbb{Z}_2^m\rightarrow\mathbb{Z}_q$ be a GBF of the variables $x_0,x_1,\hdots,x_{m-1}$.
Let us assume that $p$ denotes a prime number and define the following PBFs with the help of the GBF $f$:      
\begin{equation}\label{psb1}
\begin{split}
 F^\lambda&=f+\frac{\lambda q}{p}(x_{m}+2x_{m+1}+\cdots+2^{s-1}x_{m+s-1}),\\
 G^\lambda&=\tilde{f}+\frac{\lambda q}{p}(x_{m}+2x_{m+1}+\cdots+2^{s-1}x_{m+s-1}),
 \end{split}
\end{equation}
where $s\in \mathbb{Z}^+$ which denotes the set of all positive integers, $2\leq p< 2^{s+1}$, and $\lambda=0,1,\hdots,p-1$.
From (\ref{psb1}), it is clear that $F^\lambda$ and $G^\lambda$ are PBFs of $m+s$ variables $x_0,x_1,\hdots,x_{m+s-1}$. From (\ref{psb1}), it 
can be observed that the PBFs $F^\lambda$ and $G^\lambda$ reduce to $\mathbb{Z}_q$-valued GBFs if $p$ divides $q$.
\section{Proposed Construction of Z-Complementary Code Set}
In this section, we shall present our proposed construction of ZCCS using PBFs. To this end, we first present a lemma
which will be used in our proposed construction.
\begin{lemma}{(\cite{rs})}\label{lm5}
 Let $\lambda$ and $\lambda'$ be two non-negative integers, where $0\leq \lambda\neq \lambda'<p$, $p$ is a prime number as defined in Section-II. 
 Then 
 $\displaystyle\sum_{\alpha=0}^{p-1}\omega_p^{(\lambda-\lambda')\alpha}=0.$
\end{lemma}
For $0\leq t<2^k$ and $0\leq \lambda< p$, we define the following sets of PBFs:
\begin{equation}\label{thpset1}
\begin{split}
 U_t^\lambda\!\!=\!\!\Big\{F^\lambda\!\!+\!\frac{q}{2}\Big((\mathbf{d}+\mathbf{t})\cdot \mathbf{x}
 \!+\!dx_{\gamma}\Big): 
 \mathbf{d}\in \{0,1\}^k,d\in\{0,1\} \Big\},
 \end{split}
\end{equation}
and 
\begin{equation}\label{thpset2}
 V_t^\lambda\!\!=\!\!\Big\{G^\lambda\!+\!
 \frac{q}{2}\Big((\mathbf{d}+\mathbf{t})\cdot \bar{\mathbf{x}}
 \!+\!\bar{d}x_{\gamma}\Big): \mathbf{d}\in \{0,1\}^k,d\in\{0,1\}\Big\}.
\end{equation}
Let us assume that $f^{\mathbf{d},\mathbf{t},d}\!\!=\!\!f+\frac{q}{2}((\mathbf{d}+\mathbf{t})\cdot \mathbf{x}
 \!+dx_{\gamma}),
g^{\mathbf{d},\mathbf{t},d}\!\!=\!\!\tilde{f}+\frac{q}{2}((\mathbf{d}+\mathbf{t})\cdot \bar{\mathbf{x}}
 \!+\bar{d}x_{\gamma})$, in \textit{Lemma \ref{lemma3}}. We also assume $F^{\mathbf{d},\mathbf{t},d,\lambda}=F^\lambda\!\!+\!\frac{q}{2}((\mathbf{d}+\mathbf{t})\cdot \mathbf{x})
 \!+\!dx_{\gamma}$, in (\ref{thpset1}), and $G^{\mathbf{d},\mathbf{t},d,\lambda}=G^\lambda\!+\!
 \frac{q}{2}((\mathbf{d}+\mathbf{t})\cdot \bar{\mathbf{x}}
 \!+\!\bar{d}x_{\gamma})$, in (\ref{thpset2}). As per our assumption, for any choice of $\mathbf{d},\mathbf{t}\in \{0,1\}^k$, and $d\in\{0,1\}$, 
 the functions $f^{\mathbf{d},\mathbf{t},d}$ and $g^{\mathbf{d},\mathbf{t},d}$ are $\mathbb{Z}_q$-valued GBFs of $m$ variables. For any choice 
 of $\mathbf{d},\mathbf{t}\in \{0,1\}^k, d\in\{0,1\}$, and $\lambda\in\{0,1,\hdots,p-1\}$, the functions  
 $F^{\mathbf{d},\mathbf{t},d,\lambda}$ and $G^{\mathbf{d},\mathbf{t},d,\lambda}$ are PBFs of $m+s$ variables.
We define $\psi(F^{\mathbf{d},\mathbf{t},d,\lambda})$, the complex-valued sequence corresponding to $F^{\mathbf{d},\mathbf{t},d,\lambda}$, as
\begin{equation}\label{thpolyiy3}
 \psi(F^{\mathbf{d},\mathbf{t},d,\lambda})=
 (\omega^{F^{\mathbf{d},\mathbf{t},d,\lambda}_0}_q, \omega^{F^{\mathbf{d},\mathbf{t},d,\lambda}_1}_q,\hdots, \omega_q^{F^{\mathbf{d},\mathbf{t},d,\lambda}_{2^{m+s}-1}}),
 \end{equation}
 where $F^{\mathbf{d},\mathbf{t},d,\lambda}_{r'}=F^{\mathbf{d},\mathbf{t},d,\lambda}(r_0,r_1,\cdots,r_{m+s-1})$, $r'=\displaystyle\sum_{\alpha=0}^{m+s-1}r_\alpha 2^\alpha$. 
The $r'$-th component of $\psi(F^{\mathbf{d},\mathbf{t},d,\lambda})$ is given by
\begin{equation}\label{wvvd}
 \begin{split}
\omega^{F^{\mathbf{d},\mathbf{t},d,\lambda}}_q&=\omega_q^{f^{\mathbf{d},\mathbf{t},d}(r_0,r_1,\hdots,r_{m-1})+\frac{q\lambda }{p}(r_{m}+2r_{m+1}+\cdots+2^{s-1}r_{m+s-1})}\\
&=\omega_q^{f^{\mathbf{d},\mathbf{t},d}_{r}}\omega_p^{\lambda(r_{m}+2r_{m+1}+\cdots+2^{s-1}r_{m+s-1})}.
 \end{split}
\end{equation}
 From (\ref{wvvd}), it can be observed that $\omega^{F^{\mathbf{d},\mathbf{t},d,\lambda}_{r'}}_q$ is a root of the polynomial: $z^{\delta}-1$, where
 $\delta=lcm(p,q)$, denotes a positive integer given by the least common multiple (lcm) of $p$ and $q$. Therefore, the components of 
$\psi(F^{\mathbf{d},\mathbf{t},d,\lambda})$ are given by the roots of the polynomial: $z^{\delta}-1$. From (\ref{thpolyiy3}) and (\ref{wvvd}), we have
 \begin{equation}\label{thpoly3}
 \begin{split}
 &\psi(F^{\mathbf{d},\mathbf{t},d,\lambda})
 =(\underbrace{\omega_q^{f^{\mathbf{d},\mathbf{t},d}_0}\omega_p^{\lambda (0)},\omega_q^{f^{\mathbf{d},\mathbf{t},d}_1}\omega_p^{\lambda (0)},\hdots,\omega_q^{f^{\mathbf{d},\mathbf{t},d}_{2^m-1}}\omega_p^{\lambda (0)}},\\&
 \underbrace{\omega_q^{f^{\mathbf{d},\mathbf{t},d}_0}\omega_p^{\lambda (1)},\omega_q^{f^{\mathbf{d},\mathbf{t},d}_1}\omega_p^{\lambda (1)},\hdots,\omega_q^{f^{\mathbf{d},\mathbf{t},d}_{2^m-1}}\omega_p^{\lambda (1)}},\hdots \\&
 \underbrace{\omega_q^{f^{\mathbf{d},\mathbf{t},d}_0}\omega_p^{\lambda (2^s-1)},\omega_q^{f^{\mathbf{d},\mathbf{t},d}_1}\omega_p^{\lambda(2^s-1)},\hdots,\omega_q^{f^{\mathbf{d},\mathbf{t},d}_{2^m-1}}\omega_p^{\lambda(2^s-1)}}).
 \end{split}
\end{equation}  
Let us also define $\psi_{2^{m+s}-p2^m}(F^{\mathbf{d},\mathbf{t},d,\lambda})$ which is defined to be obtained from $\psi(F^{\mathbf{d},\mathbf{t},d,\lambda})$ by removing
its last $2^{m+s}-p2^m$ components.
\begin{equation}\label{thpoly33}
 \begin{split}
 &\psi_{2^{m+s}-p2^m}(F^{\mathbf{d},\mathbf{t},d,\lambda})\\
=&(\underbrace{\omega_q^{f^{\mathbf{d},\mathbf{t},d}_0}\omega_p^{\lambda (0)},\omega_q^{f^{\mathbf{d},\mathbf{t},d}_1}\omega_p^{\lambda (0)},\hdots,\omega_q^{f^{\mathbf{d},\mathbf{t},d}_{2^m-1}}\omega_p^{\lambda (0)}},\\&
 \underbrace{\omega_q^{f^{\mathbf{d},\mathbf{t},d}_0}\omega_p^{\lambda (1)},\omega_q^{f^{\mathbf{d},\mathbf{t},d}_1}\omega_p^{\lambda (1)},\hdots,\omega_q^{f^{\mathbf{d},\mathbf{t},d}_{2^m-1}}\omega_p^{\lambda (1)}},\hdots \\&
 \underbrace{\omega_q^{f^{\mathbf{d},\mathbf{t},d}_0}\omega_p^{\lambda (p-1)},\omega_q^{f^{\mathbf{d},\mathbf{t},d}_1}\omega_p^{\lambda(p-1)},\hdots,\omega_q^{f^{\mathbf{d},\mathbf{t},d}_{2^m-1}}\omega_p^{\lambda(p-1)}}).
 \end{split}
\end{equation}
Similarly, we can also obtain
$\psi_{2^{m+s}-p2^m}(G^{\mathbf{d},\mathbf{t},d,\lambda})$ as
\begin{equation}\label{thpoly4}
 \begin{split}
 &\psi_{2^{m+s}-p2^m}(G^{\mathbf{d},\mathbf{t},d,\lambda})\\
=&(\underbrace{\omega_q^{g^{\mathbf{d},\mathbf{t},d}_0}\omega_p^{\lambda (0)},\omega_q^{g^{\mathbf{d},\mathbf{t},d}_1}\omega_p^{\lambda (0)},\hdots,\omega_q^{g^{\mathbf{d},\mathbf{t},d}_{2^m-1}}\omega_p^{\lambda (0)}},\\&
 \underbrace{\omega_q^{g^{\mathbf{d},\mathbf{t},d}_0}\omega_p^{\lambda (1)},\omega_q^{g^{\mathbf{d},\mathbf{t},d}_1}\omega_p^{\lambda (1)},\hdots,\omega_q^{g^{\mathbf{d},\mathbf{t},d}_{2^m-1}}\omega_p^{\lambda (1)}},\hdots \\&
 \underbrace{\omega_q^{g^{\mathbf{d},\mathbf{t},d}_0}\omega_p^{\lambda (p-1)},\omega_q^{g^{\mathbf{d},\mathbf{t},d}_1}\omega_p^{\lambda(p-1)},\hdots,\omega_q^{g^{\mathbf{d},\mathbf{t},d}_{2^m-1}}\omega_p^{\lambda(p-1)}}).
 \end{split}
\end{equation}
\begin{theorem}
 Let $f:\mathbb{Z}_2^m\rightarrow \mathbb{Z}_q^m$ be a GBF as defined in \textit{Lemma \ref{lemma3}}. Then the set of codes
 \begin{equation}
 \left\{\psi_{2^{m+s}-p2^m}(U_t^\lambda), \psi^*_{2^{m+s}-p2^m}(V_t^\lambda):0\leq t<2^k,0\leq \lambda<p\right\},
 \end{equation}
forms $(p2^{k+1},2^m)$-ZCCS$_{2^{k+1}}^{p2^m}$.
\end{theorem}
\begin{IEEEproof}
In (\ref{thpoly3}), (\ref{thpoly33}), and (\ref{thpoly4}), each of the parentheses below a complex-valued sequence contains $2^m$ components of the
complex-valued sequence. It can be observed that the $2^m$ components in the $i$-th parentheses of $\psi_{2^{m+s}-p2^m}(F^{\mathbf{d},\mathbf{t},d,\lambda})$ and
$\psi_{2^{m+s}-p2^m}(G^{\mathbf{d},\mathbf{t},d,\lambda})$ represent the complex-valued sequences $\omega_p^{\lambda (i-1)}\psi(f^{\mathbf{d},\mathbf{t}})$ and
$\omega_p^{\lambda (i-1)}\psi(g^{\mathbf{d},\mathbf{t}})$, respectively, where $i=1,2,\hdots,p$.
Using (\ref{thpset1}), (\ref{thpoly33}), \textit{Lemma \ref{lemma3}}, and \textit{Lemma \ref{lm5}}, the ACCF between $\psi_{2^{m+s}-p2^m}(U_t^\lambda)$ 
and $\psi_{2^{m+s}-p2^m}(U_{t'}^{\lambda '})$ for $\tau=0$ can be derived as follows:
\begin{equation}\label{thpoly5}
 \begin{split}
  \theta&(\psi_{2^{m+s}-p2^m}(U_t^\lambda),\psi_{2^{m+s}-p2^m}(U_{t'}^{\lambda '}))(0)\\
  &=\displaystyle\sum_{\mathbf{d},d}\theta(\psi_{2^{m+s}-p2^m}(F^{\mathbf{d},\mathbf{t},d,\lambda}),\psi_{2^{m+s}-p2^m}(F^{\mathbf{d},\mathbf{t}',d,\lambda'}))(0)
  \\&=\sum_{\mathbf{d},d}\theta(\psi(f^{\mathbf{d},\mathbf{t},d}),\psi(f^{\mathbf{d},\mathbf{t}',d}))(0)\sum_{\alpha=0}^{p-1}\omega_p^{(\lambda-\lambda')\alpha}\\
  &=\theta(\psi(C_t),\psi(C_{t'}))(0)\displaystyle\sum_{\alpha=0}^{p-1}\omega_p^{(\lambda-\lambda')\alpha}\\
  &=\begin{cases}
     p2^{m+k+1},& t=t',\lambda=\lambda',\\
     0         ,& t=t', \lambda\neq\lambda',\\
     0         ,& t\neq t',\lambda=\lambda',\\
     0         ,& t\neq t',\lambda\neq \lambda'.
    \end{cases}
 \end{split}
\end{equation}
Again, Using (\ref{thpset1}), (\ref{thpoly33}), and \textit{Lemma \ref{lemma3}}, the ACCF between $\psi_{2^{m+s}-p2^m}(U_t^\lambda)$ 
and $\psi_{2^{m+s}-p2^m}(U_{t'}^{\lambda '})$ for $0<|\tau|<2^m$ can be derived as
\begin{equation}\label{thpoly6}
 \begin{split}
  \theta&(\psi_{2^{m+s}-p2^m}(U_t^\lambda),\psi_{2^{m+s}-p2^m}(U_{t'}^{\lambda '}))(\tau)\\=&\theta(\psi(C_t),\psi(C_{t'}))(\tau)\displaystyle\sum_{\alpha=0}^{p-1}\omega_p^{(\lambda-\lambda')\alpha}
  \\&+\theta(\psi(C_t),\psi(C_{t'}))(\tau-2^m)\displaystyle\sum_{\alpha=0}^{p-2}\omega_p^{\lambda(\alpha+1)-\lambda'\alpha}.
 \end{split}
\end{equation}
From \textit{Lemma \ref{lemma3}}, we have 
\begin{equation}\label{thpoly7}
\begin{split}
 \theta(\psi(C_t),\psi(C_{t'}))(\tau)=0,~ 0<|\tau|<2^m.
 \end{split}
\end{equation}
From (\ref{thpoly6}) and (\ref{thpoly7}), we have 
\begin{equation}\label{thpoly8}
 \begin{split}
    \theta(\psi_{2^{m+s}-p2^m}(U_t^\lambda),\psi_{2^{m+s}-p2^m}(U_{t'}^{\lambda '}))(\tau)\!=\!0,0<|\tau|<2^m.
 \end{split}
\end{equation}
From (\ref{thpoly5}) and (\ref{thpoly8}), we have 
\begin{equation}\label{thpoly9}
 \begin{split}
  \theta&(\psi_{2^{m+s}-p2^m}(U_t^\lambda),\psi_{2^{m+s}-p2^m}(U_{t'}^{\lambda '}))(\tau)\\
  &=\begin{cases}
     p2^{m+k+1},& t=t',\lambda=\lambda',\tau=0,\\
     0         ,& t=t', \lambda\neq\lambda',0<|\tau|<2^m,\\
     0         ,& t\neq t',\lambda=\lambda',0<|\tau|<2^m,\\
     0         ,& t\neq t',\lambda\neq \lambda',0<|\tau|<2^m.
    \end{cases}
 \end{split}
\end{equation}
Similarly, it can be shown that
\begin{equation}\label{thpoly9}
 \begin{split}
  \theta&(\psi^*_{2^{m+s}-p2^m}(V_t^\lambda),\psi^*_{2^{m+s}-p2^m}(V_{t'}^{\lambda '}))(\tau)\\
  &=\begin{cases}
     p2^{m+k+1},& t=t',\lambda=\lambda',\tau=0,\\
     0         ,& t=t', \lambda\neq\lambda',0<|\tau|<2^m,\\
     0         ,& t\neq t',\lambda=\lambda',0<|\tau|<2^m,\\
     0         ,& t\neq t',\lambda\neq \lambda',0<|\tau|<2^m.
    \end{cases}
 \end{split}
\end{equation}
From \textit{Lemma \ref{lemma3}}, (\ref{thpset1}), (\ref{thpset2}), (\ref{thpoly33}), and (\ref{thpoly4}), the ACCF between $\psi_{2^{m+s}-p2^m}(U_t^\lambda)$ 
and $\psi^*_{2^{m+s}-p2^m}(V_{t'}^{\lambda '})$ for $\tau=0$ can be derived as 
\begin{equation}\label{thpoly10}
 \begin{split}
  \theta&(\psi_{2^{m+s}-p2^m}(U_t^\lambda),\psi^*_{2^{m+s}-p2^m}(V_{t'}^{\lambda '}))(0)\\&=\theta(\psi(C_t),\psi^*(\bar{C}_{t'}))(0)\displaystyle\sum_{\alpha=0}^{p-1}\omega_p^{(\lambda+\lambda')\alpha}.
 \end{split}
\end{equation}
From \textit{Lemma \ref{lemma3}}, we have
\begin{equation}\label{thpoly11}
 \theta(\psi(C_t),\psi^*(\bar{C}_{t'}))(0)=0.
\end{equation}
From (\ref{thpoly10}) and (\ref{thpoly11}), we have
\begin{equation}\label{thpoly12}
 \theta(\psi_{2^{m+s}-p2^m}(U_t^\lambda),\psi^*_{2^{m+s}-p2^m}(V_{t'}^{\lambda '}))(0)=0.
\end{equation}
From \textit{Lemma \ref{lemma3}}, (\ref{thpset1}), (\ref{thpset2}), (\ref{thpoly33}), (\ref{thpoly4}), and (\ref{thpoly11}), the ACCF between $\psi_{2^{m+s}-p2^m}(U_t^\lambda)$ 
and $\psi^*_{2^{m+s}-p2^m}(V_{t'}^{\lambda '})$ for $0<|\tau|<2^m$ can be derived as 
\begin{equation}\label{thpoly13}
 \begin{split}
  \theta&(\psi_{2^{m+s}-p2^m}(U_t^\lambda),\psi^*_{2^{m+s}-p2^m}(V_{t'}^{\lambda '}))(\tau)\\=&\theta(\psi(C_t),\psi^*(\bar{C}_{t'}))(\tau)\displaystyle\sum_{\alpha=0}^{p-1}\omega_p^{(\lambda+\lambda')\alpha}
  \\&+\theta(\psi(C_t),\psi^*(\bar{C}_{t'}))(\tau-2^m)\displaystyle\sum_{\alpha=0}^{p-2}\omega_p^{\lambda(\alpha+1)+\lambda'\alpha}\\
  =&0.
 \end{split}
\end{equation}
From (\ref{thpoly12}) and (\ref{thpoly13}), we have 
\begin{equation}\label{thpoly14}
 \theta(\psi_{2^{m+s}-p2^m}(U_t^\lambda),\psi^*_{2^{m+s}-p2^m}(V_{t'}^{\lambda '}))(\tau)=0,~|\tau|<2^m.
\end{equation}
The obtained results in (\ref{thpoly8}), (\ref{thpoly9}), and (\ref{thpoly14}) show that the following set of codes 
$$\left\{\psi_{2^{m+s}-p2^m}(U_t^\lambda),\psi^*_{2^{m+s}-p2^m}(V_t^\lambda):0\leq t<2^k,0\leq \lambda<p\right\}$$
forms $(p2^{k+1},2^m)$-ZCCS$_{2^{k+1}}^{p2^m}$.
\end{IEEEproof}
The proposed $(p2^{k+1},2^m)$-ZCCS$_{2^{k+1}}^{p2^m}$ is optimal as it satisfies the equality $K=M\lfloor\frac{N}{Z}\rfloor$. 
\begin{remark}
 For $p=2$, $\delta=lcm(p,q)=q$, and the PBFs $F^\lambda$ and $G^\lambda$ become GBFs of $m+s$ variables over $\mathbb{Z}_q$. For the same value of $p$, 
from \textit{Theorem 1}, we obtain $(2^{k+2},2^m)$-ZCCS$_{2^{k+1}}^{2^{m+1}}$ which is optimal and the components of each codewords from a code in $(2^{k+2},2^m)$-ZCCS$_{2^{k+1}}^{2^{m+1}}$ 
are drawn from the roots of the polynomial: $z^q-1$.
Therefore, the proposed construction also generates ZCCSs of length in the form of power-of-two over the ring $\mathbb{Z}_q$.
\end{remark}
Let us illustrate the Theorem 1 with the following 
example:
\begin{example}
 Let us assume that $q=2$, $p=3$, $m=3$, $k=1$ and $s=2$. Let us take the GBF $f:\{0,1\}^3\rightarrow\mathbb{Z}_2$ as follows:
 $f=x_1x_2$, where $G(f\arrowvert_{x_0=0})$ and $G(f\arrowvert_{x_0=1})$ give a path with $x_2$ as one of the end vertices. From (\ref{psb1}), we have
 \begin{equation}
  \begin{split}
  F^\lambda=x_1x_2+\frac{2\lambda }{3}(x_3+2x_4),~
   G^\lambda=\bar{x}_1\bar{x}_2+\frac{2\lambda}{3}(x_3+2x_4),
  \end{split}
 \end{equation}
where $\lambda=0,1,2$. From (\ref{thpset1}) and (\ref{thpset2}), we have
\begin{equation}
\begin{split}
 U_t^\lambda&=\left\{F^\lambda+d_0x_0+t_0x_0+dx_2:d_0,d\in\{0,1\}\right\}\\
 V_t^\lambda&=\left\{G^\lambda+d_0\bar{x}_0+t_0\bar{x}_0+\bar{d}x_2:d_0,d\in\{0,1\}\right\},
 \end{split}
\end{equation}
where $(t_0)$ is the binary vector representation of $t$. Therefore, $\{\psi_8(U_t^\lambda),\psi_8^*(V_t^\lambda):0\leq t\leq 1, 0\leq \lambda\leq 2\}$
forms $(12,8)$-ZCCS$_{4}^{24}$ which also optimal. The components of each code word from 
a code in $(12,8)$-ZCCS$_{4}^{24}$ are drawn from the roots of the polynomial: $z^\delta-1$, where $\delta\!=\! lcm(p,q)\!=\! lcm(2,3)\!\!=\!6$.
\end{example}
\begin{remark}
From (\ref{thpoly33}) and (\ref{thpoly4}), we see that $\psi_{2^{m+s}-p2^m}(U_t^\lambda)$ and $\psi^*_{2^{m+s}-p2^m}(V_t^\lambda)$
can also be expressed as the concatenation of $\omega_p^{\lambda(i-1)}\psi(C_t)$ and $\omega_p^{-\lambda(i-1)}\psi^*(\bar{C}_t)$, respectively, where
$i=1,2,\hdots,p$.
Therefore, the proposed PBF generators establish a link between the proposed direct construction and the indirect constructions 
of ZCCSs which are obtained by performing cocatenation operation on the CCCs from \cite{arthina}.
\end{remark}
\section{Conclusions}
In this paper, we have developed a direct construction of optimal ZCCS with NPT lengths.
Unlike the current state-of-the-art which can only generate sub-optimal ZCCSs with NPT lengths, the novelty of this work stems 
from the use of  PBFs.
\bibliographystyle{IEEEtran}
\bibliography{IEEE_SPL_2021}

\begin{thebibliography}{10}
\providecommand{\url}[1]{#1}
\csname url@samestyle\endcsname
\providecommand{\newblock}{\relax}
\providecommand{\bibinfo}[2]{#2}
\providecommand{\BIBentrySTDinterwordspacing}{\spaceskip=0pt\relax}
\providecommand{\BIBentryALTinterwordstretchfactor}{4}
\providecommand{\BIBentryALTinterwordspacing}{\spaceskip=\fontdimen2\font plus
\BIBentryALTinterwordstretchfactor\fontdimen3\font minus
  \fontdimen4\font\relax}
\providecommand{\BIBforeignlanguage}[2]{{%
\expandafter\ifx\csname l@#1\endcsname\relax
\typeout{** WARNING: IEEEtran.bst: No hyphenation pattern has been}%
\typeout{** loaded for the language `#1'. Using the pattern for}%
\typeout{** the default language instead.}%
\else
\language=\csname l@#1\endcsname
\fi
#2}}
\providecommand{\BIBdecl}{\relax}
\BIBdecl

\bibitem{chen2007next}
H.-H. Chen, \emph{The Next Generation CDMA Technologies}.\hskip 1em plus 0.5em
  minus 0.4em\relax Wiley, 2007.

\bibitem{mai1}
D.~{Carey}, D.~{Roviras}, and B.~{Senadji}, ``Comparison of multiple access
  interference in asynchronous {MC-CDMA} and {DS-CDMA} systems,'' in
  \emph{Proceedings Seventh International Symposium on Signal Processing and
  Its Applications.}, vol.~2, 2003, pp. 351--354.

\bibitem{mpi1}
P.~{Nagaradjane}, A.~{Swaminathan}, K.~S. {Dhyaneshwaran}, B.~R.
  {Narayanasamy}, and A.~{Ramakrishnan}, ``Multipath interference mitigation
  technique for {MC DS/CDMA} systems,'' in \emph{International Conference on
  Control, Automation, Communication and Energy Conservation}, 2009, pp. 1--3.

\bibitem{arthina}
A.~Rathinakumar and A.~K. Chaturvedi, ``Complete mutually orthogonal {G}olay
  complementary sets from {Reed-Muller} codes,'' \emph{IEEE Trans. Inf.
  Theory}, vol.~54, no.~3, pp. 1339--1346, Mar. 2008.

\bibitem{psktcom}
P.~{Sarkar}, S.~{Majhi}, and Z.~{Liu}, ``Optimal {Z} -complementary code set
  from generalized {Reed-Muller} codes,'' \emph{IEEE Trans. Commun}, vol.~67,
  no.~3, pp. 1783--1796, Mar. 2019.

\bibitem{fan2007}
P.~Fan, W.~Yuan, and Y.~Tu, ``Z-complementary binary sequences,'' \emph{IEEE
  Signal Process. Lett.}, vol.~14, no.~8, pp. 509--512, Aug. 2007.

\bibitem{gol1961}
M.~Golay, ``Complementary series,'' \emph{IRE Trans. Inf. Theory}, vol.~7,
  no.~2, pp. 82--87, Apr. 1961.

\bibitem{Davis1999}
J.~A. Davis and J.~Jedwab, ``Peak-to-mean power control in {OFDM}, {Golay}
  complementary sequences, and {Reed-Muller} codes,'' \emph{IEEE Trans. Inf.
  Theory}, vol.~45, no.~7, pp. 2397--2417, Nov. 1999.

\bibitem{chen}
C.~{Chen}, ``A novel construction of {Z}-complementary pairs based on
  generalized {Boolean} functions,'' \emph{IEEE Signal Process. Lett.},
  vol.~24, no.~7, pp. 987--990, July 2017.

\bibitem{czcp3}
C.~{Pai}, S.~{Wu}, and C.~{Chen}, ``{Z}-complementary pairs with flexible
  lengths from generalized {Boolean} functions,'' \emph{IEEE Commun. Lett.},
  vol.~24, no.~6, pp. 1183--1187, 2020.

\bibitem{wespl}
A.~R. {Adhikary}, P.~{Sarkar}, and S.~{Majhi}, ``A direct construction of
  $q$-ary even length {Z}-complementary pairs using generalized {Boolean}
  functions,'' \emph{IEEE Signal Process. Lett.}, vol.~27, pp. 146--150, 2020.

\bibitem{lfengispl2008}
L.~Feng, P.~Fan, X.~Tang, and K.~K. Loo, ``Generalized pairwise
  {Z}-complementary codes,'' \emph{IEEE Signal Process. Lett.}, vol.~15, pp.
  377--380, 2008.

\bibitem{liu2011}
Z.~Liu, Y.~L. Guan, B.~C. Ng, and H.-H. Chen, ``Correlation and set size bounds
  of complementary sequences with low correlation zone,'' \emph{IEEE Trans.
  Commun.}, vol.~59, no.~12, pp. 3285--3289, Dec. 2011.

\bibitem{pater2000}
K.~G. Paterson, ``Generalized {Reed-Muller} codes and power control in {OFDM}
  modulation,'' \emph{IEEE Trans. Inf. Theory}, vol.~46, no.~1, pp. 104--120,
  Jan. 2000.

\bibitem{isit19}
P.~{Sarkar}, S.~{Majhi}, and Z.~{Liu}, ``A direct and generalized construction
  of polyphase complementary set with low {PMEPR},'' in \emph{2019 IEEE
  International Symposium on Information Theory (ISIT)}, 2019, pp. 2279--2283.

\bibitem{tcom20}
------, ``A direct and generalized construction of polyphase complementary sets
  with low {PMEPR} and high code-rate for {OFDM} system,'' \emph{IEEE Trans.
  Commun.}, 2020.

\bibitem{cccsmajhi}
S.~Das, S.~Budi\v{s}in, S.~Majhi, Z.~Liu, and Y.~L. Guan, ``A multiplier-free
  generator for polyphase complete complementary codes,'' \emph{IEEE Trans.
  Signal Process.}, vol.~66, no.~5, pp. 1184--1196, Mar. 2018.

\bibitem{uda2014}
Z.~Liu, Y.~L. Guan, and U.~Parampalli, ``New complete complementary codes for
  peak-to-mean power control in multi-carrier {CDMA},'' \emph{IEEE Trans.
  Commun.}, vol.~62, no.~3, pp. 1105--1113, Mar. 2014.

\bibitem{Liu_FDRR_2015}
Z.~Liu, Y.~L. Guan, and H.-H. Chen, ``Fractional-delay-resilient receiver
  design for interference-free {MC-CDMA} communications based on complete
  complementary codes,'' \emph{IEEE Trans. Wireless Commun.}, vol.~14, no.~3,
  pp. 1226--1236, Mar. 2015.

\bibitem{slett}
S.~Das, S.~Majhi, and Z.~Liu, ``A novel class of complete complementary codes
  and their applications for apu matrices,'' \emph{IEEE Signal Process. Lett.},
  vol.~25, no.~9, pp. 1300--1304, Sept. 2018.

\bibitem{ps2cd}
P.~{Sarkar} and S.~{Majhi}, ``A direct construction of optimal zccs with
  maximum column sequence {PMEPR} two for {MC-CDMA} system,'' \emph{IEEE
  Commun. Lett.}, 2020.

\bibitem{chuzccs}
S.~W. {Wu}, A.~{Şahin}, Z.~M. {Huang}, and C.~Y. {Chen}, ``{Z}-complementary
  code sets with flexible lengths from generalized {Boolean} functions,''
  \emph{IEEE Access}, vol.~9, pp. 4642--4652, 2021.

\bibitem{chenzcom}
S.~{Wu} and C.~{Chen}, ``Optimal {Z}-complementary sequence sets with good
  peak-to-average power-ratio property,'' \emph{IEEE Signal Process. Lett.},
  vol.~25, no.~10, pp. 1500--1504, Oct. 2018.

\bibitem{palnpl}
P.~{Sarkar}, A.~{Roy}, and S.~{Majhi}, ``Construction of {Z}-complementary code
  sets with non-power-of-two lengths based on generalized {Boolean}
  functions,'' \emph{IEEE Commun. Lett.}, pp. 1--5, 2020.

\bibitem{tibudazccs}
S.~{Das}, U.~{Parampalli}, S.~{Majhi}, Z.~{Liu}, and S.~{Budišin}, ``New
  optimal {Z}-complementary code sets based on generalized paraunitary
  matrices,'' \emph{IEEE Trans. Signal Process.}, vol.~68, pp. 5546--5558,
  2020.

\bibitem{avezccs}
A.~Adhikary and S.~Majhi, ``New construction of optimal aperiodic
  {Z}-complementary sequence sets of odd-lengths,'' \emph{Electron. Lett.},
  vol.~55, no.~19, pp. 1043--1045, 2019.

\bibitem{jli_igc_2008}
J.~Li, A.~Huang, M.~Guizani, and H.-H. Chen, ``Inter group complementary codes
  for interference resistant {CDMA} wireless communications,'' \emph{IEEE
  Trans. Wireless Commun.}, vol.~7, no.~1, pp. 166--174, Jan. 2008.

\bibitem{fan2008}
W.~Yuan, Y.~Tu, and P.~Fan, ``Optimal training sequences for
  cyclic-prefix-based single-carrier multi-antenna systems with space-time
  block-coding,'' \emph{IEEE Trans. Wireless Commun.}, vol.~7, no.~11, pp.
  4047--4050, Nov. 2008.

\bibitem{hmwang2007}
H.~M. Wang, X.~Q. Gao, B.~Jiang, X.~H. You, and W.~Hong, ``Efficient {MIMO}
  channel estimation using complementary sequences,'' \emph{IET Commun.},
  vol.~1, no.~5, pp. 962--969, Oct. 2007.

\bibitem{psbb}
V.~K. {Leont’ev}, ``On pseudo-boolean polynomials,'' \emph{Comput. Math. and
  Math. Phys.}, vol.~55, pp. 1926--1932, 2015.

\bibitem{rs}
P.~P. {Vaidyanathan}, ``{Ramanujan} sums in the context of signal
  processing-{Part I: Fundamentals},'' \emph{IEEE Trans. Signal Process.},
  vol.~62, no.~16, pp. 4145--4157, 2014.

\end{thebibliography}
\end{document}